\documentclass[conference]{IEEEtran}
\IEEEoverridecommandlockouts
\usepackage{color, soul}
\usepackage{float}

\usepackage{cite}
\usepackage{amsmath,amssymb,amsfonts}
\usepackage{algorithmic}
\usepackage{graphicx}
\usepackage{textcomp}
\usepackage{xcolor}
\usepackage{braket}
\usepackage{balance}
\usepackage[a4paper, total={184mm,239mm}]{geometry}

\newcommand\numberthis{\addtocounter{equation}{1}\tag{\theequation}}

\begin{document}

\title{ArsoNISQ: Analyzing Quantum Algorithms on Near-Term Architectures
}

\author{
\thanks{This work was partially funded by the Carl Zeiss foundation.}

\begin{tabular}{@{}c@{\qquad\quad}c@{}}
\multicolumn{2}{c}{Sebastian Brandhofer$^{1}$ \qquad Simon Devitt$^2$ \qquad Ilia Polian$^{1}$}\\[0.5cm]
    \begin{tabular}{@{}c@{}}
\normalsize
    $^1$Institute of Computer Architecture and Computer Engineering,\\
\normalsize
    Center for Integrated Quantum Science and Technology (IQ$^\text{ST}$)\\
\normalsize
    University of Stuttgart, Stuttgart, Germany \\
\normalsize
    \{sebastian.brandhofer\;$|$\;ilia.polian\}@informatik.uni-stuttgart.de
    \end{tabular}
&
    \begin{tabular}{@{}c@{}}
\normalsize
    $^2$Centre for Quantum Software and Information \\
\normalsize
    University of Technology Sydney, Sydney, Australia \\
\normalsize
    simon.devitt@uts.edu.au\hspace{0cm}\\
    \\
    \end{tabular}

\end{tabular}
\vspace{-0.5ex}
}

\maketitle

\begin{abstract}
While scalable, fully error corrected quantum computing is years or even decades away, there is considerable interest in noisy intermediate-scale quantum computing (NISQ).
In this paper, we introduce the \texttt{ArsoNISQ} framework that determines the tolerable error rate of a given quantum algorithm computation, i.e. quantum circuits, and the success probability of the computation given a success criterion and a NISQ computer.
\texttt{ArsoNISQ} is based on simulations of quantum circuits subject to errors according to the Pauli error model.

\texttt{ArsoNISQ} was evaluated on a set of quantum algorithms that can incur a quantum speedup or are otherwise relevant to NISQ computing.
Despite optimistic expectations in recent literature, we did not observe quantum algorithms with intrinsic robustness, i.e. algorithms that tolerate one error on average, in this evaluation.
The evaluation demonstrated, however, that the quantum circuit size sets an upper bound for its tolerable error rate and quantified the difference in tolerate error rates for quantum circuits of similar sizes.
Thus, the framework can assist quantum algorithm developers in improving their implementation and selecting a suitable NISQ computing platform.
Extrapolating the results into the quantum advantage regime suggests that the error rate of larger quantum computers must decrease substantially or active quantum error correction will need to be deployed for most of the evaluated algorithms.

\end{abstract}

\begin{IEEEkeywords}
Quantum Computing, NISQ Computing, Error Simulation, Error Tolerance Analysis, Quantum Algorithm Design
\end{IEEEkeywords}
\section{Introduction}

Quantum computing promises an exponential speed up for problems in cryptography~\cite{39}, chemistry and material science~\cite{13, 24}. 
For example, approximately $3\%$ of the world's energy supply are currently consumed by industrial processes for fertilizer production~\cite{23}.
A quantum computer capable of error-free operation can perform classically intractable simulations that yield data for reducing the energy requirements of these processes~\cite{0}.
However, current quantum computers cannot deliver such error-free operation, and they also do not fulfill the conditions for effective quantum error correction schemes~\cite{26}.

The recent progress in quantum technology has spurred interest in \emph{noisy and intermediate-scale quantum} (NISQ) computing, loosely referring to systems with between 50 and a few hundred quantum bits (qubits) and an error rate exceeding 0.1\% per operation~\cite{22}. These systems can be based on different technologies such as superconducting circuits~\cite{11, 41, 28}, ion traps~\cite{38,37,12} or others. Among the reasons for the widespread engagement for---and investment into---NISQ computing is the recent demonstration of an existing NISQ computer that took 3.3 minutes~\cite{11}~for a problem that requires immense classical resources~\cite{20}. This has led to expectations that NISQ computers are only a short step ahead of outperforming their classical counterparts for practically useful problems, even without deploying quantum error correction protocols~\cite{11, 22}.

At the same time, due to the large impact of incorrectable errors, it is poorly understood which quantum algorithms can be computed successfully on NISQ computers~\cite{9, 15, 22}.
However, this is crucial for selecting a suitable NISQ computer that facilitates successful computation of an algorithm, improving the implementation of a quantum algorithm and deciding whether quantum error correction protocols are required.

The work at hand addresses these aspects by developing and evaluating the novel framework \texttt{ArsoNISQ}:
\begin{itemize}
    \item \texttt{ArsoNISQ} is a systematic technique applicable to arbitrary quantum circuits and gate-based NISQ computers.
    \item \texttt{ArsoNISQ} reduces the simulation effort for quantum circuits subject to noise.
    \item The evaluation of \texttt{ArsoNISQ} establishes and quantifies the relationship between the size of a quantum computation and its tolerable error rate.
    \item Extrapolation of \texttt{ArsoNISQ} results allow statements about the quantum advantage regime, i.e. quantum circuits that can not be simulated classically.
\end{itemize}

The \texttt{ArsoNISQ} framework ``puts a quantum circuit, in combination with a NISQ computer, on fire'' and computes to what extent this combination can be expected to produce useful results.
This expectation is quantified as the tolerable error rate of a quantum circuits and as the success probability of such a computation on a specified NISQ computer.

The remainder of this work is structured as follows.
In section~\ref{5}, quantum computing fundamentals are described.
In section~\ref{2}~the related work is reviewed.
The \texttt{ArsoNISQ} framework is described in section~\ref{12} and exemplary results are shown in section~\ref{9}.
Finally, we conclude this work in section~\ref{4}.

\section{Quantum Computing}\label{5}
Quantum computing deals with the preparation, manipulation, storage, communication and measurement of quantum states.
The quantum state of the basic unit of quantum information, i.e. one qubit, is represented by:
\begin{equation}
    \ket{\psi} = \alpha_{0} \ket{0} + \alpha_{1} \ket{1}
\end{equation}
where $\ket{0}, \ket{1}$ are basis states and $\alpha_k$ are complex probability amplitudes.
The quantum state of $n$ qubits is fully described by $2^n$ complex probability amplitudes corresponding to $2^n$ basis states.
Measuring an $n$-qubit quantum state $\ket{\psi}$ yields a classical value $k$ corresponding to the basis state $\ket{k}$ with probability $|\alpha_{k}|^2$ (where $\alpha_k$ corresponds to $\ket{k}$).

\begin{figure}[t!]
  \centering
  \includegraphics[width=\linewidth]{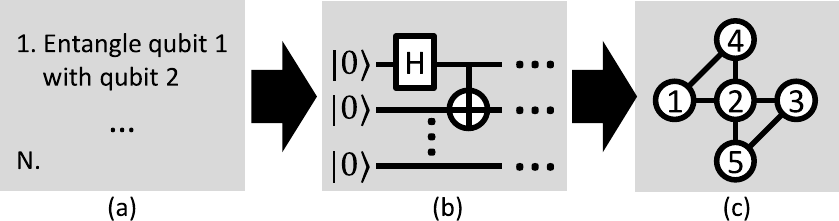}       \vspace{-4ex}
  \caption{Mapping of a quantum algorithm (a) to a quantum circuit (b) that is computed on a quantum computer subject to geometric constraints (c).\label{6}}
  \vspace{-2ex}
\end{figure}

An $n$-qubit quantum computer is capable of acting on an $n$-qubit quantum state depending on external controls.
Current quantum computers have been realized based on superconducting circuits~\cite{11, 41, 28}, ion traps~\cite{38,37,12}, point defects in diamonds (NV centers)~\cite{47,43}~or photons~\cite{32,27}.
Only specific quantum state initializations, manipulations and measurements are supported by a quantum computer depending on their physical realization.
For instance, for quantum computers based on superconducting circuits, only neighboring physically connected qubits can interact with each other~\cite{11, 41}.
These geometric constraints are represented by a connectivity graph such as depicted in figure~\ref{6}~(c).

A quantum algorithm consists of abstract steps describing the preparation of one or multiple quantum states that yield the solution to a problem upon measurement.
Quantum algorithms such as Shor's~\cite{39}~or Grover's~\cite{30}~algorithm incur an asymptotic runtime improvement compared to classical algorithms for certain problems.

Before a computation on a quantum computer, a quantum algorithm must be implemented as a quantum circuit that satisfies the constraints given by the physical hardware.
A quantum circuit consists of quantum gates that specify how the probability amplitudes of a quantum state are manipulated.
Figure~\ref{6}~shows the implementation of one abstract step in a quantum algorithm as a quantum circuit with one single-qubit Hadamard gate and one two-qubit controlled-not gate~\cite{33}.
That quantum circuit is then computed on a quantum computer that exhibits geometric constraints depicted by a connectivity graph.
In this work, the size of a quantum circuit is defined as the number of its quantum gates.

A state vector simulator computes a $n$-qubit quantum circuit by storing the $2^n$ complex amplitudes of the quantum state classically up to some precision and manipulating the amplitudes as specified by the quantum gates in the quantum circuit.
As such, a general state vector simulator incurs an exponential runtime in the number of qubits, which is the lower runtime bound of simulators that can accurately execute arbitrary quantum circuits~\cite{33}.
In general, a density matrix simulator stores and manipulates $2^n$ $n$-qubit quantum states for a $n$-qubit quantum circuit computation, i.e. $2^{2n}$ complex amplitudes~\cite{33}.

The success probability of computing a quantum circuit can be assessed using different success criteria~\cite{6}.
When the probability amplitudes of a quantum state are known, a typical criterion is the fidelity $F(\ket{\psi}, \ket{\psi'})$ that quantifies how much an erroneous state $\ket{\psi'}$ deviates from the ideal state $\ket{\psi}$.
When only measurement results are available, other criteria are used to compare states $\ket{\psi}$ and $\ket{\psi'}$ including:
\begin{itemize}
    \item The probability of measuring the correct result.
    \item The probability of a measurement being in a defined set of acceptable results (binning).
    \item The cross-entropy of measurement results~\cite{25}.
\end{itemize}
Fidelity, measurement probability and binning is used in the evaluation of this work.

In the Pauli error model, Pauli $X, Z$ or $Y$ errors, corresponding to bit-, phase- or bitphase-flips, can occur each time a quantum gate acts on a quantum state.
The Pauli error rate quantifies the percentage of gates that are followed by such a Pauli error.

After a single-qubit gate, a Pauli $X, Z$ or $Y$ error would transform the state $\ket{\psi} = a_{0}\ket{0} + a_{1}\ket{1}$ to $\ket{\psi_{X}}, \ket{\psi_{Z}}$ or $\ket{\psi_{Y}}$ with
\begin{alignat*}{3}
        \ket{\psi_{X}}  &= &&a_{1}\ket{0}  + &&a_{0} \ket{1} \\
        \ket{\psi_{Z}}  &= &&a_{0}\ket{0} -&& a_{1} \ket{1} \\
        \ket{\psi_{Y}}  &= -i\cdot &&a_{1}\ket{0} + i\cdot &&a_{0} \ket{1}.  
        \numberthis\label{eqn}
\end{alignat*}
A combination of Pauli errors and the identity may occur after a two-qubit gate, in the Pauli error model.
There are 16 such combinations and we only exemplarily show the effect of two Pauli $X$ errors on the respective qubits in $\ket{\psi} = a_{0}\ket{00} + a_{1}\ket{01} + a_{2}\ket{10} + a_{3}\ket{11}$ as $\ket{\psi_{XX}}=a_{3}\ket{00} + a_{2}\ket{01} + a_{1}\ket{10} + a_{0}\ket{11}$.

\section{Related Work}\label{2}

\texttt{ArsoNISQ} determines both, the tolerable error rate of a quantum circuit and the success probability of executing such a circuit on a given NISQ computer for a specified success criterion.
To the best of the authors' knowledge, there is no previous systematic technique that quantitatively determines the tolerable error rate of quantum circuits for sizes indicative for the NISQ era.
In~\cite{22}~and~\cite{9}, the relationship between quantum computation size and tolerable error rate is described qualitatively, while the work at hand demonstrates the upper bound of this relationship and quantifies it.
In~\cite{22}, it is postulated that the tolerable error rate of a successful quantum computation on a NISQ computer may not be much larger than $G^{-1}$, where $G$ is the size of that computation given in number of quantum gates.
In~\cite{9}, an opposite bound between the error rate and the size of a quantum computation is conjectured where the error rate must be much lower than $(n\cdot d)^{-1}$ with $n$ being the number of qubits and $d$ being the depth of the quantum computation.

In related work, the success of a quantum circuit execution is assessed by executing the quantum circuit on a target quantum computer, performing error simulation, using estimates based on the the size of the quantum circuit~\cite{22, 9}~or consulting performance metrics such as the quantum volume~\cite{17}.
While the success of a quantum circuit execution can be assessed by running it on a physical quantum computer, this assessment can not be generalized to other quantum computing technologies or even future generations of the same quantum computer architecture~\cite{2}.
\texttt{ArsoNISQ} addresses this aspect by employing an error model that is applicable to diverse NISQ computers and technologies.

Previous error simulation works assessing the success of a quantum computation focus on the impact of noise on one particular quantum algorithm~\cite{3, 19, 10, 4, 21, 16, 18}~or investigate \emph{device-oriented} error models~\cite{8, 18, 10, 3, 36}.
Furthermore, previous exhaustive error simulations~\cite{19,16,18, 40,29,36}~employ the density operator formalism that requires more memory and a much larger runtime, effectively halving the number of qubits of analyzable quantum circuits compared to the state vector simulator~\cite{33}~employed by \texttt{ArsoNISQ}~\cite{4}.
In addition, \texttt{ArsoNISQ} is flexible with respect to the NISQ algorithm and the success criterion and generates a detailed relation between between success probability and quantum computation size.

\section{The ArsoNISQ Framework}\label{12}
\texttt{ArsoNISQ} determines the success probability and tolerable Pauli error rate of a quantum algorithm computation.
\texttt{ArsoNISQ} assumes a quantum algorithm computation, i.e. a quantum circuit, and a success criterion as input.
In addition, either a Pauli error rate or a success probability must be provided.
If a target success probability is provided, the tolerable Pauli error rate is computed.
The tolerable error rate is the maximal error rate a quantum circuit can be subject to while still matching the success probability of a given success criterion.
If instead a Pauli error rate is provided, \texttt{ArsoNISQ} computes the success probability of a quantum circuit execution subject to the specified error rate.

Computing the success probability and the tolerable error rate of a quantum algorithm requires different approaches that are described in section~\ref{11}~and section~\ref{13} respectively.
Both of these computations are based on simulations subject to Pauli errors, which will be explained in section~\ref{8}~after justifying the choice of the Pauli error model~in section~\ref{0}.

\subsection{Error Modeling}\label{0}

Errors in a quantum computation are modeled by replicating adverse physical processes (\emph{noise processes}) affecting a quantum computer~\cite{33}~or by selecting a set of operations that cover the impact of relevant noise sources on a quantum computation~\cite{1, 34}.
We call the former a device-oriented and the latter a device-agnostic error model.
A device-oriented error model replicates noise processes leading to crosstalk errors~\cite{7}, systematic errors, qubit loss~\cite{1}, or other errors that are specific to one quantum computer of a particular technology.
The Pauli error model covers the effect of coherent and incoherent noise stemming from measurement, initialization and other quantum state manipulations excluding qubit loss and leakage~\cite{1}.

Determining the success probability and tolerable error rate of a quantum circuit poses various requirements on an error model:
\begin{enumerate}
    \item The error model must be accurate enough to allow predictions about the success of a quantum computation.
    \item The error model must be efficiently quantifiable as a physical error rate of a NISQ computer in principle.
    \item The error model must be sufficiently general to be applicable to a wide range of NISQ computers and technologies to support the selection of a NISQ computer.
    \item The error model should not increase the simulation effort substantially.
\end{enumerate}
In this work, the device-agnostic Pauli error model is used as it satisfies above requirements:
\begin{enumerate}
    \item Experiments in~\cite{11}~validated that the noise processes in a state of the art NISQ computer based on superconducting circuits are represented accurately by the Pauli error model, i.e. noise processes do not need to be replicated.

    \item Protocols applicable to arbitrary quantum computers efficiently and accurately quantify the Pauli error model as an Pauli error rate~\cite{14, 11}.
    \item A device-oriented error model of a NISQ computer is not applicable to other quantum computing technologies and may not be indicative for future NISQ computer generations of the same technology~\cite{2}.
    On the other side, the Pauli error model is ubiquitous in many quantum computing technologies~\cite{14}, which enables comparison over these technologies.
    \item Replicating noise processes requires a density matrix simulator in general whose memory requirement halves the size of simulatable quantum circuits compared to state vector simulators applicable the Pauli error model.
\end{enumerate}

\subsection{Pauli Error Simulation}\label{8}

The error simulations of \texttt{ArsoNISQ} are conducted using a state vector simulator that computes the exact impact of an error on the state yielded by the given quantum circuit~\cite{4}.
This allows the framework to evaluate a success criterion subject to a specific Pauli error accurately in a subsequent step.
The given quantum circuit is simulated subject to a Pauli error after a quantum gate by appending an error gate to that quantum gate to the quantum circuit simulation.
This error gate represents the probability amplitude changes incurred by the error (see section~\ref{5}).

To lower the simulation effort of arbitrary pairs of quantum circuits and error rates, two simulation approaches are employed.
If the quantum circuit is subject to a low number of errors on average, a large number of Monte Carlo simulation runs are required to gather accurate statistics about the impact of these errors.
For such simulation instances, \texttt{ArsoNISQ} performs an exhaustive Pauli error simulation and scales its outcome.
However, the number of error combinations evaluated in an exhaustive error simulation increases exponentially with the number of considered errors in the quantum circuit.
Therefore, for simulation instances with more than one error per quantum circuit on average, a Monte Carlo simulation is conducted.

\subsubsection{Exhaustive Pauli Error Simulation}

The reference state $\ket{\psi_{R}}$ is first computed by simulating the quantum circuit without errors.
Then, the quantum circuit is successively simulated subject to every potential single Pauli $X, Z,$ and $Y$ error at every quantum gate in the quantum circuit, which yields the state set
\begin{equation}
\left\{\ket{\psi_{X_{1}}}\!,\! ..,\! \ket{\psi_{X_{k+2m}}}\!,\! \ket{\psi_{Z_{1}}}\!,\! ..,\! \ket{\psi_{Z_{k+2m}}}\!,\! \ket{\psi_{Y_{1}}}\!,\! ..,\! \ket{\psi_{Y_{k+2m}}}\!\right\}
\end{equation}
in a circuit with $k$ single-qubit gates and $m$ two-qubit gates.

\subsubsection{Monte Carlo Error Simulation}
In a Monte Carlo error simulation the quantum circuit is simulated $N$ times subject to the Pauli error model.
After a single-qubit gate is executed in a quantum circuit simulation, a Pauli $X, Z$ or $Y$ error occurs with probabilities $p_{X},p_{Z},p_{Y}$.
For two-qubit gates, a Pauli $X, Z$ or $Y$ error occurs on each gate qubit independently with the same probabilities.
In our experiments, we observed good accuracy for $N=1000$ simulations per quantum circuit and error rate.

\subsection{Success Probability}\label{11}
\begin{figure}[t!]
  \vspace{-1ex}
  \centering
  \includegraphics[width=0.85\linewidth]{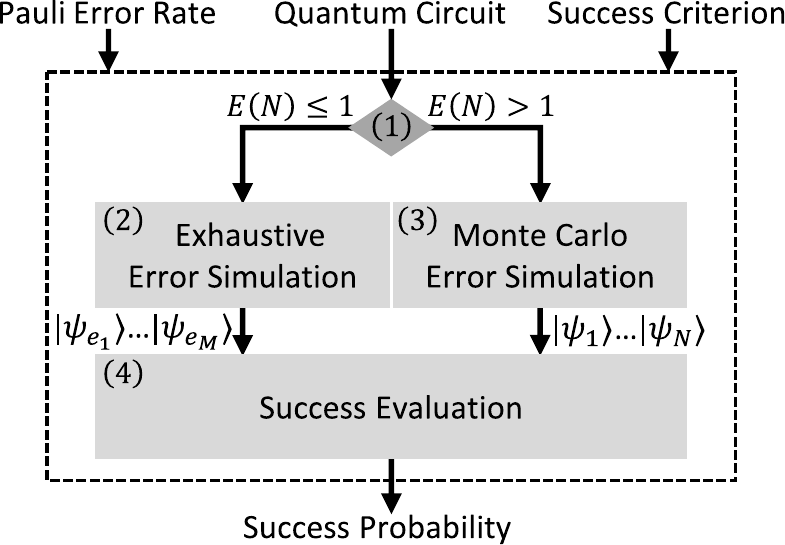}
  
  \vspace{-2ex}
   \caption{Steps performed by \texttt{ArsoNISQ} to determine the success probability given a quantum circuit, success criterion and Pauli error rate. Depending on condition (1), either exhaustive (2) or Monte Carlo (3) error simulation is performed and the outcomes undergo success criterion evaluation (4)\label{3}}
  
 \vspace{-2ex}
\end{figure}

The steps in figure~\ref{3}~depict how the success probability of a quantum algorithm is determined.
First, the average number of Pauli errors $E(N)$ is computed from the given error rate and quantum circuit.
If the average number of errors is larger than one, a Monte Carlo error simulation is conducted. 
Otherwise, we exhaustively map out the effect of a single error at all possible circuit locations.
On the result of these simulations, the success probability is evaluated based on the specified error rate and success criterion.

The success probability is evaluated differently depending on the preceding error simulation.
If an exhaustive error simulation was conducted, the success probability $P_\psi$ of the specified success criterion is evaluated on every erroneous state $\ket{\psi_{e_{i}}}$ and the reference state $\ket{\psi_{R}}$.
The success probability is then averaged over all states that are subject to the same type of Pauli error.
This yields success probabilities $P_{\psi_{X}}, P_{\psi_{Z}}$ and $P_{\psi_{Y}}$ for each type of error.
The average number of Pauli $X$, $Z$ and $Y$ errors $E(N_{X})$, $E(N_{Z})$ and $E(N_{Y})$ is computed as the product of the number of quantum gates in the quantum circuit and the Pauli error rates $p_{X}, p_{Z}, p_{Y}$ respectively. 

The average success probability $P$ of a quantum circuit subject to single Pauli errors can then be computed as
\begin{equation}\label{1}
    P = \!\!\sum_{e\in \{X, Z, Y\}} \!\!P_{\psi_{e}}E(N_{e}) \!+ P_{\psi_{R}}\!\cdot\! \left(\!1 - \!\sum_{e\in \{X, Z, Y\}} E(N_{e})\!\right)
\end{equation}
when the average number of errors is at most one.
Otherwise, if a Monte Carlo error simulation was conducted, the average success $P$ is determined by averaging the probability $P_\psi$ of the specified success criterion over every simulated state $\psi_{1},...,\psi_N$.

\subsection{Tolerable Pauli Error Rate}\label{13}
\begin{figure}[t!]
  \centering

  \includegraphics[width=0.72\linewidth]{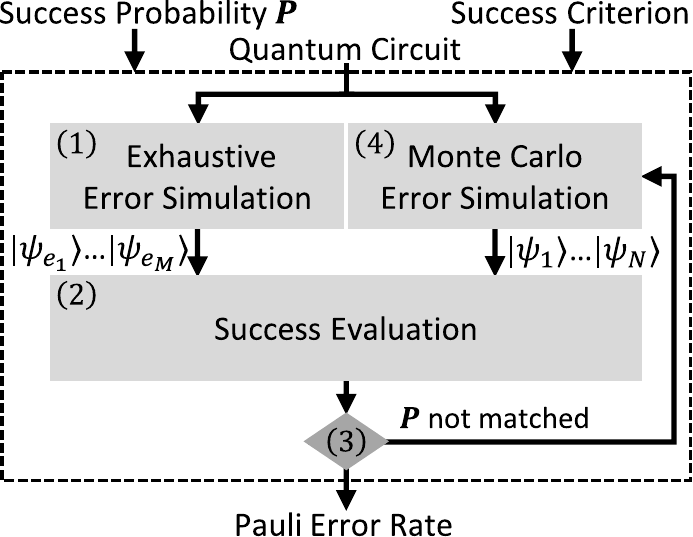}
    \vspace{-1ex}
  \caption{Steps performed by \texttt{ArsoNISQ} to determine the tolerable Pauli error rate given a quantum circuit, success criterion and success probability. Exhaustive error simulation (1) is applied to the quantum circuit, and as long as success evaluation (2) remains above the target success probability $P$ (3), Monte Carlo simulation with increasing error rates (4) is performed.\label{7}}
    
    \vspace{-2ex}
\end{figure}

The steps in figure~\ref{7}~depict how the tolerable Pauli error rate is determined given a quantum circuit, a success probability and a success criterion.
First, an exhaustive error simulation is conducted on the given quantum circuit.
This yields a state set that represents the impact of each single error on the quantum circuit execution.
The average success probability of the quantum circuit subject to one average error is evaluated as shown in the previous section.

If the success probability evaluation indicates that at most one error is tolerable, the tolerable uniform error rate $p$ $(p/3=p_{X}=p_{Z}=p_{Y})$ given a target success probability $P$ can be derived from equation~\ref{1} as
\begin{equation}\label{15}
p  =  G^{-1} \cdot \frac{P - P_{\psi_{R}}}{\sum_{e\in \{X, Z, Y\}} \frac{1}{3}P_{\psi_{e}} -P_{\psi_{R}}}
\end{equation}
where $G$ is the number of gates in the quantum circuit and $E(N_{e})=G \cdot p_{e} = G  \cdot p/3$.
From equation~\ref{15}~it follows directly that the tolerable error rate $p$ is smaller than $G^{-1}$,
if the target success probability $P$ is larger than the success probability subject to one average error $\sum_{e\in \{X, Z, Y\}} \frac{1}{3}P_{\psi_{e}}$.

If the success probability evaluation indicates that more than one error can be tolerated, Monte Carlo error simulations with increasingly higher error rates are conducted until the given target success probability is matched.
Powell's optimization method~\cite{31}~is used to converge to a error rate that matches the given success probability.

\section{Evaluation}\label{9}
In this section, \texttt{ArsoNISQ} is evaluated on a set of quantum circuits.
For these circuits, the success probability and tolerable error rate is reported.
The evaluated set of quantum circuits implement arithmetic functions~\cite{42}, the Grover algorithm~\cite{30}, quantum Fourier transform (QFT)~\cite{44}, the hidden linear function problem (HLF)~\cite{45} and the Bernstein-Vazirani (BV) algorithm~\cite{46} (using controlled-not gates).
In addition, quantum 'ansatz' circuits for simulating molecules (UCCSD~\cite{35}~and RYRZ~\cite{5}) and quantum volume (QV)~\cite{17}~circuits are evaluated.
Depending on the quantum circuit, one of three success criteria was evaluated.
The measurement probability is reported for quantum circuits with one single correct outcome, for the other quantum circuits excluding QV circuits, the fidelity measure was chosen.
For QV circuits a form of binning was used, i.e. the quantum volume success probability (heavy output probability) as defined in~\cite{17}~was evaluated and averaged over 200 random quantum volume circuits for each data point.

The evaluated quantum algorithms were mapped to quantum circuits adapted to a general quantum computer without geometric constraints using Qiskit~\cite{29}.
Imposing geometric constraints on the quantum algorithm will decrease the tolerable error rate further.
Thus, subsequent results present an upper bound on the success probability and the tolerable error rate. 
However, \texttt{ArsoNISQ} is applicable to quantum algorithms mapped to any geometric constraint.
Quantum circuits with up to 16 qubits and 1066 gates were evaluated.
An uniform error rate was assumed, i.e. a Pauli $X, Z$ or $Y$ error occurs independently with the same probability.
The evaluation was conducted on Intel Core i7 machines with at least 16 GB RAM.
On these machines and on the evaluated 4-, 10-, 12-, 14- and 16-qubit quantum circuits, ArsoNISQ incurred a runtime of 0.06 seconds, 6.84 seconds, 66.4 seconds, 10.68 minutes and 16.6 minutes respectively per quantum circuit, on average.

\subsection{Success Probability}\label{91}
\begin{figure}[t!]
  \centering
  \includegraphics[width=\linewidth]{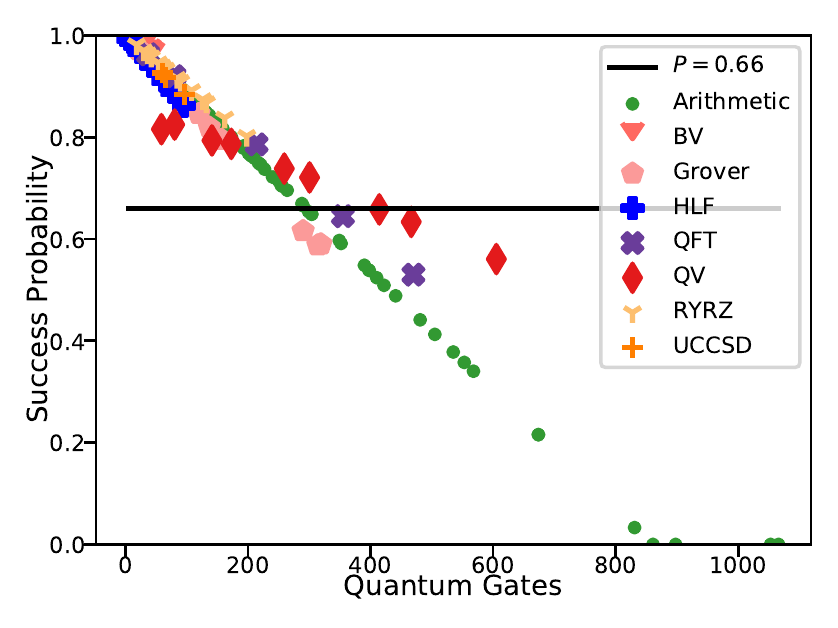}
  
  \caption{Success probability of quantum circuits subject to a Pauli error rate of 0.15\% and no geometric constraints. \label{10}}
  \vspace{-3ex}

\end{figure}

In figure~\ref{10}~the success probability of quantum circuit execution subject to a Pauli error rate of $0.15\%$ is depicted.
This rate is the current lowest Pauli error rate reported for superconducting qubits~\cite{11}.
The success probability is shown on the y-axis and the number of quantum gates in a quantum circuit is shown on the x-axis.
The results show that the success probability of the evaluated quantum circuits
decreases linearly in the size of their corresponding quantum circuits at a fixed error rate.
All evaluated quantum circuits with less than 300 gates could be executed successfully.
The largest evaluated quantum circuits such as the 16-qubit QFT quantum circuit and the 14-qubit Grover quantum circuit could not be executed successfully.
QV circuits that indicate a quantum volume of 100 were on the verge of being successfully executed.

The results also show that quantum circuits with similar size can have large differences in success probability at the same error rate as evident when comparing arithmetic quantum circuits to QFT or QV quantum circuits in figure~\ref{10}.
Furthermore, the combination of QV circuits and their success criterion consistently exhibited the largest success probability for quantum circuits with more than 260 gates.

Thus, if a quantum algorithm developer is given a quantum computer with known Pauli error rate and various quantum algorithm implementations, \texttt{ArsoNISQ} assists in selecting the quantum circuit with the highest probability of success and estimating the largest executable quantum circuit.

\subsection{Pauli Error Rate}\label{92}

\begin{figure}[t!]
  \centering
  \includegraphics[width=\linewidth]{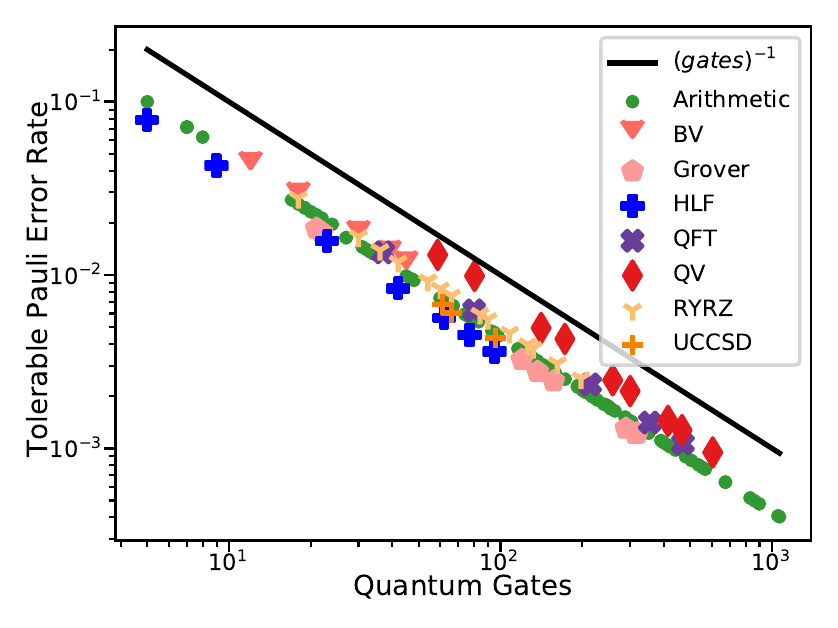}

  \caption{Tolerable Pauli error rate of the evaluated quantum circuits for a success probability of 66\%.\label{14}}
   \vspace{-3.6ex}
   
\end{figure}

Figure~\ref{14}~shows the tolerable error rate (y-axis) of the evaluated quantum circuits for a success probability of 66\%.
The number $G$ of quantum gates in a quantum circuit is shown on the x-axis.
A logarithmic scale was used for both axes.

The tolerable error rate of all evaluated quantum algorithms was upper bounded by $G^{-1}$, i.e. less than one error on average was tolerated.
This upper bound was up to $2.9\times$ $(2.22\times$ on average$)$ as large as the tolerable error rate computed by \texttt{ArsoNISQ}.
With the set of evaluated arithmetic quantum circuits being the average, QV, BV, QFT and RYRZ quantum circuits tolerate a slightly larger error rate while HLF, Grover and UCCSD quantum circuits could only tolerate a slightly lower error rate.
The largest difference in tolerable error rate for quantum circuits of similar size is between HLF and QV where HLF quantum circuits tolerate an error rate that is up to $45\%$ smaller than for QV quantum circuits.

Thus, \texttt{ArsoNISQ} supports a quantum algorithm developer in exploiting this difference in error susceptibility to improve the quantum circuit.
In addition, a quantum algorithm developer is assisted in selecting a suitable NISQ computer for the realization of a quantum algorithm.

\subsection{Extrapolations to the Quantum Advantage Regime}
As established in figure~\ref{14}, the tolerable error rate of each evaluated quantum algorithm class is roughly linear in $G^{-1}$.
This observation was used to extrapolate the tolerable error rate of a quantum algorithm into the quantum advantage regime, i.e. where classical simulations are considered intractable.
Quantum circuits with around 55 qubits are currently intractable to simulate accurately on a classical computer in general~\cite{11, 20}.
Therefore, we generated a 56-qubit quantum circuit and fitted a function linear in $G^{-1}$ for each evaluated quantum algorithm class in figure~\ref{14}.
The incurred mean squared error of the fitted function corresponding to the arithmetic quantum algorithms was $\approx 5.7\cdot 10^{-7}$.
The mean squared error was lower for all other algorithms.

From the evaluated 56-qubit quantum circuits, only the BV and the RYRZ quantum circuit of depth one tolerated a error rate (0.3\% and 0.151\%) that can be achieved by current NISQ computers based on superconducting circuits~\cite{11}.
The 56-qubit quantum circuits of Grover's algorithm, HLF and QFT were extrapolated to tolerate a error rate of roughly $10^{-4}$.
The lowest tolerable error rate was observed for the 56-qubit UCCSD quantum circuit $(10^{-6})$.
These extrapolation results indicate that the error rate must improve by several orders of magnitude or active quantum error correction must be employed for most of the evaluated quantum algorithms.
\texttt{ArsoNISQ} can therefore also assist a quantum algorithm developer when considering quantum circuits with sizes prohibiting classical simulation.

\section{Conclusion}\label{4}

In this work, \texttt{ArsoNISQ}, a framework for quantifying the tolerable error rate and success probability of an arbitrary quantum circuit for a specified success criterion was developed.
An evaluation on intermediate-scale quantum circuits shows that the success probability decreases linearly in the number of quantum gates $G$ and that the tolerable error rate decreases as $G^{-1}$.
This evaluation was extrapolated to the quantum advantage regime where it indicated that the error rate of current quantum computers must decrease substantially or active quantum error correction must be employed for most of the evaluated quantum algorithms.

\texttt{ArsoNISQ} assists a quantum algorithm developer in selecting a NISQ computer with suitable error characteristics, improving their quantum computation by exploiting differences in the tolerable error rate and to formulate error rate requirements for larger quantum computers of upcoming generations.
Through extrapolations, this assistance can be extended to quantum computations that are too large to be simulated classically.
\vspace{-0.2ex}

\balance
\scriptsize

\bibliographystyle{IEEEtran}

\bibliography{aspdac_nisq_compatibility3}

\end{document}